# About the standard methodology in electron-molecule collision physics

A. S. Baltenkov[1] and I. Woiciechowski[2]

[1]*Arifov Institute of Ion-Plasma and Laser Technologies,
100125, Tashkent, Uzbekistan*
[2]*Alderson Broaddus University, 101 College Hill Drive,
WV 26416, Philippi, USA*

**Abstract:** The article discusses the correctness of the assumption about the similarity of molecular continuum electron functions with wave functions in electron-atom scattering. The elastic scattering of slow particles by pair of non-overlapping short-range potentials has been studied. The continuum wave function of particle is represented as a combination of a plane wave and two spherical s-waves, generated by the scattering centers. The asymptotic of this function determines in closed form the amplitude of elastic particle scattering. It is shown that this amplitude of scattering on a nonspherical target cannot be correctly represented as an expansion into a series of spherical functions $Y_{lm}(\mathbf{r})$, as is the case for scattering on an atom. Therefore, methods of the scattering phase calculation based on the assumption of spherical symmetry of the molecular potential field beyond the molecular sphere cannot be considered as justified and reliable. So is the representation outside this sphere of the wave function as a linear combination of the regular and irregular solutions of the Schrödinger equation. It has been shown that asymptotically at great distances from the molecule the continuum wave functions can be presented as an expansion in a set of other, than spherical, orthonormal functions $Z_\lambda(\mathbf{r})$. General formulas for these functions are obtained. The coefficients of the scattering amplitude expansion to a series of these functions determine the scattering phases for the molecular system under consideration. The special features of the S-matrix method for the case of arbitrary non-spherical potentials are discussed.

## Introduction

In the molecular collision physics, the idea that *molecular continuum electron functions are similar to those for electron-atom scattering* [1], are considered as a matter-of-course and, as far as we know, they are beyond any doubt. Adaptation of muffin-tin-model of molecule consists in introducing in this model a molecular sphere that surrounds all atoms forming the molecule [2]. The electron continuum wave functions outside this sphere represented as a linear combination of the regular and irregular solutions of the Schrödinger equation for the potential that in this region "is taken to be spherical … about the molecular center" [2]. The coefficients of this linear combination are defined by the molecular phases of scattering $\eta_l(k)$. They are defined by the matching conditions of the continuum wave function on the surfaces of the atomic and molecular spheres. Hence, in the multiple scattering method [2] the solution of the problem of electron scattering by a spherically non-symmetrical potential is reduced *without any grounds* to the usual S-matrix method of the partial waves for a spherical target. Within this approach, the asymptotic form of the wave function far from the molecule (beyond the molecular sphere) is a sum of a plane wave and spherical wave emitted by the molecular center

$$\psi_{\mathbf{k}}^{+}(\mathbf{r} \to \infty) \approx e^{i\mathbf{k}\cdot\mathbf{r}} + F(\mathbf{k},\mathbf{k}')\frac{e^{ikr}}{r}. \qquad (1)$$

Function $F(\mathbf{k},\mathbf{k}')$ here is the usual elastic scattering amplitude expanded into a series of the spherical functions

$$F(\mathbf{k},\mathbf{k}') = \frac{2\pi}{ik} \sum_{l,m} (e^{2i\eta_l} - 1) Y_{lm}^*(\mathbf{k}) Y_{lm}(\mathbf{k}') . \qquad (2)$$

The purpose of our paper is to show by a simple example that formulas (1) and (2) for nonspherical targets are no longer valid, just as the very assumption about the similarity of the processes of electron scattering by atoms and molecules is also incorrect. We understand that doubts about the correctness of the standard technique described above contradict the whole "industry" of calculations of scattering processes. However, we consider it useful to pay attention to some elements of the standard methodology that too respectful traditionalism avoids analyze.

In Cohen and Fano paper [3] we read: "Diffraction phenomena should occur when electrons released within a multi-center molecular field. From Huygens' point of view, the two atoms of $N_2$ constitute two separate sources of photoelectrons. Superposition of the emission from these two sources produces an interference pattern whose properties should depend periodically on the ratio of the inter-nuclear distance to the photoelectron wavelength".

A qualitative picture of the scattering of an electron wave by a diatomic target, according to the above quote, is shown in Fig. 1(a). On the left in the figure is an electron plane wave, and on the right are two spherical waves, as follows from the Huygens-Fresnel principle. The interference of these waves creates a diffraction scattering pattern.

Figure 1(b) allows you to see how the electron wave scattering pattern is transformed after the introduction of the molecular sphere. Outside the molecular sphere, there is a single spherical wave centered at the center of the molecular sphere. Therefore, the diffraction of electron waves within the framework of the approach [2] (as any other methods that are based on single spherical wave far from target) is simply impossible, since there is no pair of spherical waves emitted by spatially separated sources. The surface of the molecular sphere is the place where the solutions of the wave equation are joined inside and outside the sphere, and the calculated molecular phases of scattering are, in fact, the phases of scattering of an electron wave on an isolated molecular sphere, and not on a pair of atomic spheres, as it should be.

The outline of our article is as follows. In Section 1, we will calculate the scattering amplitude of a particle on a pair of short-range potentials based on the scattering pattern presented in Figure 1(a). It will be shown that this amplitude can be written in a closed form, but not as an expansion in partial waves. In section 2 we will try to expand this exact scattering amplitude into a series of spherical functions. Let us show that this attempt leads to irremovable contradictions, which indicates the inapplicability of formula (2) for non-spherical targets. In Section 3, we describe the partial wave method for the scattering pattern Figure 1(a). Expansion of the scattering amplitude in a series of functions $Z_\lambda(\mathbf{k})$ is carried out in Section 4. Section 5 is Conclusions.

## 1. Elastic scattering of a particle by two short-range potentials

Based on Brueckner's article [4], we will follow how the problem of particle scattering by a pair of non-overlapping short-range potential wells is solved in nuclear physics. The continuum wave function according to Figure 1(a) is represented as a combination of a plane wave and two spherical *s*-waves, generated by short-range potentials with centers at the points $\mathbf{r} = \pm \mathbf{R}/2$

$$\psi_\mathbf{k}^+(\mathbf{r}) = e^{i\mathbf{k}\cdot\mathbf{r}} + D_1(\mathbf{k})\frac{e^{ik|\mathbf{r}+\mathbf{R}/2|}}{|\mathbf{r}+\mathbf{R}/2|} + D_2(\mathbf{k})\frac{e^{ik|\mathbf{r}-\mathbf{R}/2|}}{|\mathbf{r}-\mathbf{R}/2|} ; \text{ for } |\mathbf{r} \pm \mathbf{R}/2| > \rho. \qquad (1)$$

Here $\rho$ is radius of potential wells. Function (1) is the general solution to the wave equation outside the scatterer's spheres [4].

The so far unknown coefficients $D_1(\mathbf{k})$ and $D_1(\mathbf{k})$ in (1) are found as a result of imposing boundary conditions on trial wave function $\psi_\mathbf{k}^+(\mathbf{r})$ at the points $\mathbf{R}/2$ and $-\mathbf{R}/2$. Let's find these boundary conditions. To do this, we study the behavior of a particle in the S-state in the field of a short-range spherical potential well. Outside the well at a distance larger than the range of potential $\rho$, the particle wave function satisfies the wave equation

$$\psi_0(\mathbf{r}) \propto \frac{1}{kr}\sin(kr+\delta_0)Y_{00}(\mathbf{r}). \qquad (2)$$

Let us assume that we know the S-phase of the elastic scattering of a particle on an isolated potential well, that is, the function $\delta_o(k)$ over the entire range of the wave vector *k*. Passing to the limit $r \to 0$ in formula (2), we obtain the following boundary conditions imposed on function (1) at the centers of potential well

$$\psi_0(\mathbf{r})_{r\to 0} \propto C\left[\frac{1}{r} + k\cot\delta_0(k)\right].$$

For wells centered at points $\mathbf{r} = \pm \mathbf{R}/2$, we obtain the following boundary conditions for function (1)

$$\psi_\mathbf{k}^+(\mathbf{r})_{\mathbf{r}\to \mathbf{R}/2} \approx C_1\left[\frac{1}{|\mathbf{r}-\mathbf{R}/2|} + k\cot\delta_0(k)\right];$$

$$\psi_\mathbf{k}^+(\mathbf{r})_{\mathbf{r}\to -\mathbf{R}/2} \approx C_2\left[\frac{1}{|\mathbf{r}+\mathbf{R}/2|} + k\cot\delta_0(k)\right]. \qquad (3)$$

$C_1$ and $C_2$ here are the some constants. Applying formulas (3) to the function (1) we obtain the *exact general solution* of the Schrödinger equation outside the scatterer's spheres that describes the multiple scattering of particle by two-center target [4]. The amplitude of particle scattering by the target is obtained by considering the asymptotic behavior of the wave function (1). As result, we obtain the following *exact amplitude* of the particle multiple scattering by given two-center target [4, 5]

$$F(\mathbf{k},\mathbf{k}',\mathbf{R}) = \frac{2}{a^2-b^2}\{b\cos[(\mathbf{k}-\mathbf{k}')\cdot\frac{\mathbf{R}}{2}] - a\cos[(\mathbf{k}+\mathbf{k}')\cdot\frac{\mathbf{R}}{2}]\}. \qquad (4)$$

We not reproduced here the details, but refer the reader to the original paper [5]. As in [5], we use here the following notation: $\mathbf{k}$ and $\mathbf{k}'$ are the particle linear momentums before and after scattering, respectively; the functions $a = \exp(ikR)/R$ and $b = k(i - \cot\delta_0) = -1/f_0(k)$. Here $f_0(k)$ is the s-partial amplitude of particle elastic scattering by isolated potential well. The total cross section of particle scattering in [3] is obtained from the amplitude (4) using the optical theorem [6]

$$\sigma(\mathbf{k},\mathbf{R}) = \int \frac{d\sigma}{d\Omega_{k'}} d\Omega_{k'} = \frac{4\pi}{k}\operatorname{Im} F(\mathbf{k}=\mathbf{k}',\mathbf{R}) = \frac{4\pi}{k}\operatorname{Im}\left[\frac{b - a\cos(\mathbf{k}\cdot\mathbf{R})}{a^2-b^2}\right]. \qquad (5)$$

## 2. Expansion of the scattering amplitude (4) in spherical functions

Let us analyze whether it is possible to represent the exact amplitude (4) as a series of spherical functions, like (2). To do this, we equate the function $F(\mathbf{k},\mathbf{k}',\mathbf{R})$ from (4) to the functions $F(\mathbf{k},\mathbf{k}')$ from (2). We obtain the following equation for the molecular phase shifts $\eta_l(k)$

$$\frac{2}{a^2-b^2}\{b\cos[(\mathbf{k}-\mathbf{k}')\cdot\frac{\mathbf{R}}{2}] - a\cos[(\mathbf{k}+\mathbf{k}')\cdot\frac{\mathbf{R}}{2}]\} = \frac{2\pi}{ik}\sum_{l,m}(e^{2i\eta_l}-1)Y^*_{lm}(\mathbf{k})Y_{lm}(\mathbf{k}'). \qquad (6)$$

Let us multiply both side of this equality by $Y_{lm}(\mathbf{k})Y^*_{lm}(\mathbf{k})$ and integrating over all directions of vectors $\mathbf{k}$ and $\mathbf{k}'$. The right side of equation (6) after integration becomes $2\pi(e^{2i\eta_l}-1)/ik$. The left side after integration is reduced to evaluation of the following integrals:

$$I_{lm} = \int_0^{2\pi}\int_0^{\pi} e^{i\mathbf{k}\cdot\frac{\mathbf{R}}{2}} Y_{lm}(\mathbf{k})\sin\vartheta\, d\vartheta\, d\varphi,$$

where $\theta$ and $\varphi$ are the directional angles of the vector $\mathbf{k}$ with respect to axes $\mathbf{R}$. The integral $I_{lm}$ can be evaluated using the expansion

$$e^{i\mathbf{k}\cdot\frac{\mathbf{R}}{2}} = 4\pi\sum_{l=0}^{\infty} i^l j_l(kR/2)\sum_{-l}^{l} Y^*_{lm}(\mathbf{R})Y_{lm}(\mathbf{k})$$

where $j_l(x)$ is the spherical Bessel function [7]. Taking into account the orthogonality of the spherical harmonics we obtain

$$I_{lm} = 4\pi i^l j_l(kR/2)Y^*_{lm}(\mathbf{R})$$

Finally, equation (6) reads

$$32\pi^2 j_l^2(kR/2)|Y_{lm}(\mathbf{R})|^2 \frac{b-(-1)^l a}{a^2-b^2} = \frac{2\pi}{ik}(e^{2i\eta_l}-1).$$

It can be rewritten as

$$\cos\eta_l \sin\eta_l + i\sin^2\eta_l = 8\pi k j_l^2(kR/2)|Y_{lm}(\mathbf{R})|^2 \frac{b-(-1)^l a}{a^2-b^2} = W\frac{b-(-1)^l a}{a^2-b^2}. \quad (7)$$

where $W$ is the real function. Equating the real and imaginery parts of equation (7) we obtain the equations for *the infinite number of the molecular phases* $\eta_l(k)$ [8]:

$$\cos\eta_l \sin\eta_l = W\,\mathrm{Re}\frac{b-(-1)^l a}{a^2-b^2}\,;\quad \sin^2\eta_l = W\,\mathrm{Im}\frac{b-(-1)^l a}{a^2-b^2}.$$

Dividing the first equation by the second one we obtain the following expressions for the cotangents molecular phases $\eta_l(k)$:

$$\cot\eta_l = \frac{\mathrm{Re}[(a+b)^*]}{\mathrm{Im}[(a+b)^*]},\text{ for even } l,\quad \cot\eta_l = \frac{\mathrm{Re}[(a-b)^*]}{\mathrm{Im}[(a-b)^*]},\text{ for odd } l. \quad (8)$$

It is evident that the total cross section of scattering by two centers calculated with these phases (owing to their independence on *l*), in contrast to (5), diverges. Indeed, rewriting the sum of partial cross sections as two infinite sums, we obtain

$$\bar{\sigma}(k) = \frac{4\pi}{k^2}\sum_{l=0}^{\infty}(2l+1)\sin^2\eta_l = \frac{4\pi}{k^2}\left[\sum_{l=even}^{\infty}(2l+1)\sin^2\eta_l + \sum_{l=odd}^{\infty}(2l+1)\sin^2\eta_l\right] = \infty. \quad (9)$$

The reason for this meaningless result is a manifestation of the fact that the exact amplitude of elastic scattering by a two-center target (4), in principle, cannot be represented as a partial expansion in a series of spherical functions (2). So, the straightforward application to a target that is non-spherical of the usual S-matrix method (2) is impossible.

### 3. Method of partial waves for non-spherical targets

A molecular potential as a cluster of non-overlapping atomic potentials centered at the atomic sites is non-spherical. The solution $\psi_{\mathbf{k}}^+(\mathbf{r})$ of the Schrödinger equation with this potential is impossible to present at an arbitrary point of space as an expansion in spherical functions $Y_{lm}(\mathbf{r})$. However, asymptotically at great distances from the molecule, according to Demkov and Rudakov article [9], the continuum wave function of electron can be presented as an expansion in a set of other orthonormal functions $Z_\lambda(\mathbf{r})$ and $Z_\lambda(\mathbf{k})$:

$$\psi_{\mathbf{k}}^+(\mathbf{r}\to\infty) \approx 4\pi\sum_\lambda R_{k\lambda}(r)Z_\lambda(\mathbf{r})Z_\lambda^*(\mathbf{k}) \quad (10)$$

with the radial part of the wave function determined by the following expression

$$[R_{k\lambda}(r)Z_\lambda(\mathbf{r})]_{r\to\infty} \approx e^{i(\eta_\lambda + \frac{\pi}{2}\omega_\lambda)} \frac{1}{kr} \sin(kr - \frac{\pi}{2}\omega_\lambda + \eta_\lambda) Z_\lambda(\mathbf{r}). \qquad (11)$$

Here the index $\lambda$ numerates different partial wave functions similar to the quantum numbers $l$ and $m$ for the central field; $\omega_\lambda$ is the quantum number that is equal to the orbital moment $l$ for the spherical symmetry case; $\eta_\lambda(k)$ are the "proper molecular phases". The explicit form of functions $Z_\lambda(\mathbf{k})$ (in terminology [9] "characteristic amplitudes") depends upon the specific type of the target field, particularly on the number of atoms forming the target and on mutual disposition of the scattering centers in space, *etc*. Functions $Z_\lambda(\mathbf{k})$, like the spherical functions $Y_{lm}(\mathbf{k})$, create an orthonormal system.

The elastic scattering amplitude for a non-spherical target, according to [9], is given, in contrast to (2), by the following expression

$$F(\mathbf{k},\mathbf{k}') = \frac{2\pi}{ik} \sum_\lambda (e^{2i\eta_\lambda} - 1) Z_\lambda^*(\mathbf{k}) Z_\lambda(\mathbf{k}'). \qquad (12)$$

## 4. Expansion of the scattering amplitude (4) in a series of functions $Z_\lambda(\mathbf{k})$

According to (12), the exact amplitude (4) should be represented as a partial expansion in a series of functions $Z_\lambda(\mathbf{k})$. For a given molecular system the molecular phase shifts $\eta_\lambda(k)$ and functions $Z_\lambda(\mathbf{k})$ can be calculated explicitly [10]. Following this paper let us rewrite the scattering amplitude (4) in following form

$$F(\mathbf{k},\mathbf{k}',\mathbf{R}) = -\frac{2}{a+b}\cos(\mathbf{k}\cdot\mathbf{R}/2)\cos(\mathbf{k}'\cdot\mathbf{R}/2) + \frac{2}{a-b}\sin(\mathbf{k}\cdot\mathbf{R}/2)\sin(\mathbf{k}'\cdot\mathbf{R}/2). \qquad (13)$$

The amplitude (4) should be considered as a sum of two partial amplitudes. The first of them is written as

$$\frac{4\pi}{2ik}(e^{2i\eta_0} - 1) Z_0(\mathbf{k}) Z_0^*(\mathbf{k}') = -\frac{2}{a+b}\cos(\mathbf{k}\cdot\mathbf{R}/2)\cos(\mathbf{k}'\cdot\mathbf{R}/2). \qquad (14)$$

The second one is defined by the following expression

$$\frac{4\pi}{2ik}(e^{2i\eta_1} - 1) Z_1(\mathbf{k}) Z_1^*(\mathbf{k}') = \frac{2}{a-b}\sin(\mathbf{k}\cdot\mathbf{R}/2)\sin(\mathbf{k}'\cdot\mathbf{R}/2). \qquad (15)$$

After elementary transformations of (14) and (15), we obtain two molecular phases (rather than an infinite number, as in (8))

$$\cot\eta_0 = -\frac{qR + \cos kR}{kR + \sin kR}, \qquad \cot\eta_1 = -\frac{qR - \cos kR}{kR - \sin kR}. \qquad (16)$$

Here the wave vector $q(k) = -k\cot_0(k)$. The molecular phases $\eta_\lambda(k)$ in (16) can be classified by considering their behavior at $k \to 0$ [9]. In this limit the electron wavelength is much greater than the target size and the picture of scattering should approach that in the case of spherical symmetry. Considering the transition to this limit in the formulas (16), we obtain: $\eta_0(k \to 0) \sim k$ and $\eta_1(k \to 0) \sim k^3$. Thus, the molecular phases behave similar to the *s*- and *p*- phases in the spherically symmetrical potential, which explains the choice of their indexes.

Let us establish a relation between the molecular electron scattering length $A_m$ and the electron scattering length on the atoms *A* that form the target. These scattering lengths are determined by the following formulas [11]

$$1/A_m = [-k\cot\eta_0(k)]_{k \to 0} \text{ and } 1/A = [-k\cot\delta_0(k)]_{k \to 0}.$$

Multiplying both parts of the zero phase (16) by and passing to the limit $k \to 0$, we obtain the required relation between the scattering lengths (compare with [12])

$$\frac{1}{A_m} = \frac{R+A}{2RA}.$$

From formulas (14) and (15) we obtain two "characteristic amplitudes"

$$Z_0(\mathbf{k}) = \frac{\cos(\mathbf{k}\cdot\mathbf{R}/2)}{\sqrt{2\pi S_+}}, \qquad Z_1(\mathbf{k}) = \frac{\sin(\mathbf{k}\cdot\mathbf{R}/2)}{\sqrt{2\pi S_-}}. \qquad (17)$$

Here $S_{\pm} = 1 \pm j_0(kR)$. It is easy to demonstrate that the functions $Z_\lambda(\mathbf{k})$, like the spherical functions $Y_{lm}(\mathbf{k})$, create an orthonormal system. The functions (17) are defined by the geometrical target structure, i.e. by the direction of the molecular axis **R** in the arbitrary coordinate system, in which the electron momentum vectors before and after scattering are **k** and **k'**, respectively. The transition to the limit $k \to 0$ in formulas (17) gives instead of the functions $Z_\lambda(\mathbf{k})$ the well-known spherical functions

$$Z_0(\mathbf{k})_{k\to 0} \to \frac{1}{\sqrt{4\pi}} \equiv Y_{00}(\mathbf{k}), \qquad Z_1(\mathbf{k})_{k\to 0} \to \sqrt{\frac{3}{4\pi}}\cos\vartheta \equiv Y_{10}(\mathbf{k}). \qquad (18)$$

Here $\vartheta$ is the angle between the vector **k** and axis **R**. So, *only in the limit $k \to 0$ it becomes correct to replace a non-spherical molecular field by a spherical one*. We can follow limit transitions (18) in Figures 2 and 3 [13]. Solid lines in these figures are the limits of the functions $Z_\lambda(\mathbf{k}')$ when $k \to 0$.

## 5. Conclusions

In studies of electron-molecule scattering, the continuum electron is usually treated [1, 2] as moving in a spherically averaged molecular field [see figure 1(b)]. The wave functions describing the scattering of an electron by a polyatomic molecule outside the so-called molecular sphere are considered as a linear combination of regular and irregular solutions of the wave equation. The phase shifts of molecular wave function are defined from the matching condition for the solutions of the wave equation inside and beyond this sphere. Obviously, the introduction of a molecular sphere ("by hands", without any ground) changes the scattering pattern [compare Fig.

1(a) and (b)]. The diffraction pattern of the electron wave itself disappears, since there is only one spherical wave far from the target. Whereas terms $\sin kR/kR$ in formulas (16), obtained within the scattering pattern 1(a), there is a clear manifestation of the electron diffraction of a pair of spherical s-waves emitted by spatially separated sources. Apparently, the traditional approach based on formulas (1) and (2) and Fig. (b) as a result allows one to obtain the scattering phases of a particle on an isolated molecular sphere, rather than on a real two-center target.

Thus, the statement [1] (page 361) that the *molecular continuum electron functions are similar to those for electron-atom scattering* should be supplemented with the words: *if the Huygens-Fresnel principle is neglected.*

Since in the asymptotics of the wave function of the continuum for non-spherical targets (11) instead of spherical harmonics there are functions $Z_\lambda(\mathbf{r})$, it becomes clear that: *no matter how far we move away from the non-spherical target, we will never be able to consider the target to be spherically symmetric, and the wave function of the particle to be a spherical wave*; except the case when the particle wavelength is much more than the target size $1/k >> R$.

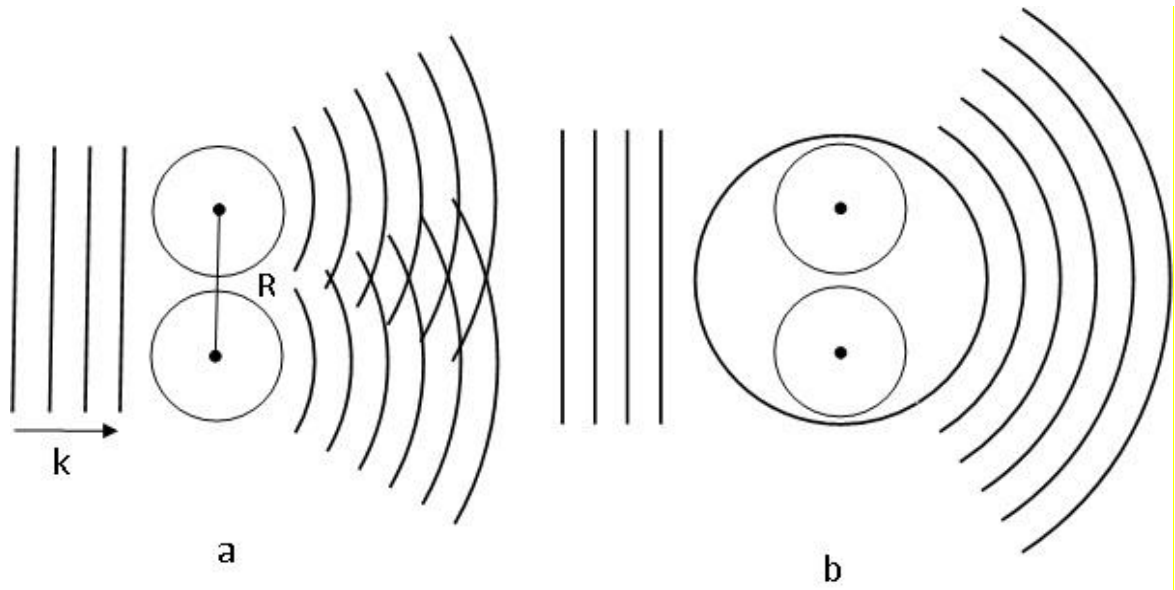

Figure 1. A qualitative picture of the scattering of an electron wave by a diatomic target; (a) is the Huygens-Fresnel picture; (b) is the Dill and Dehmer picture [2].

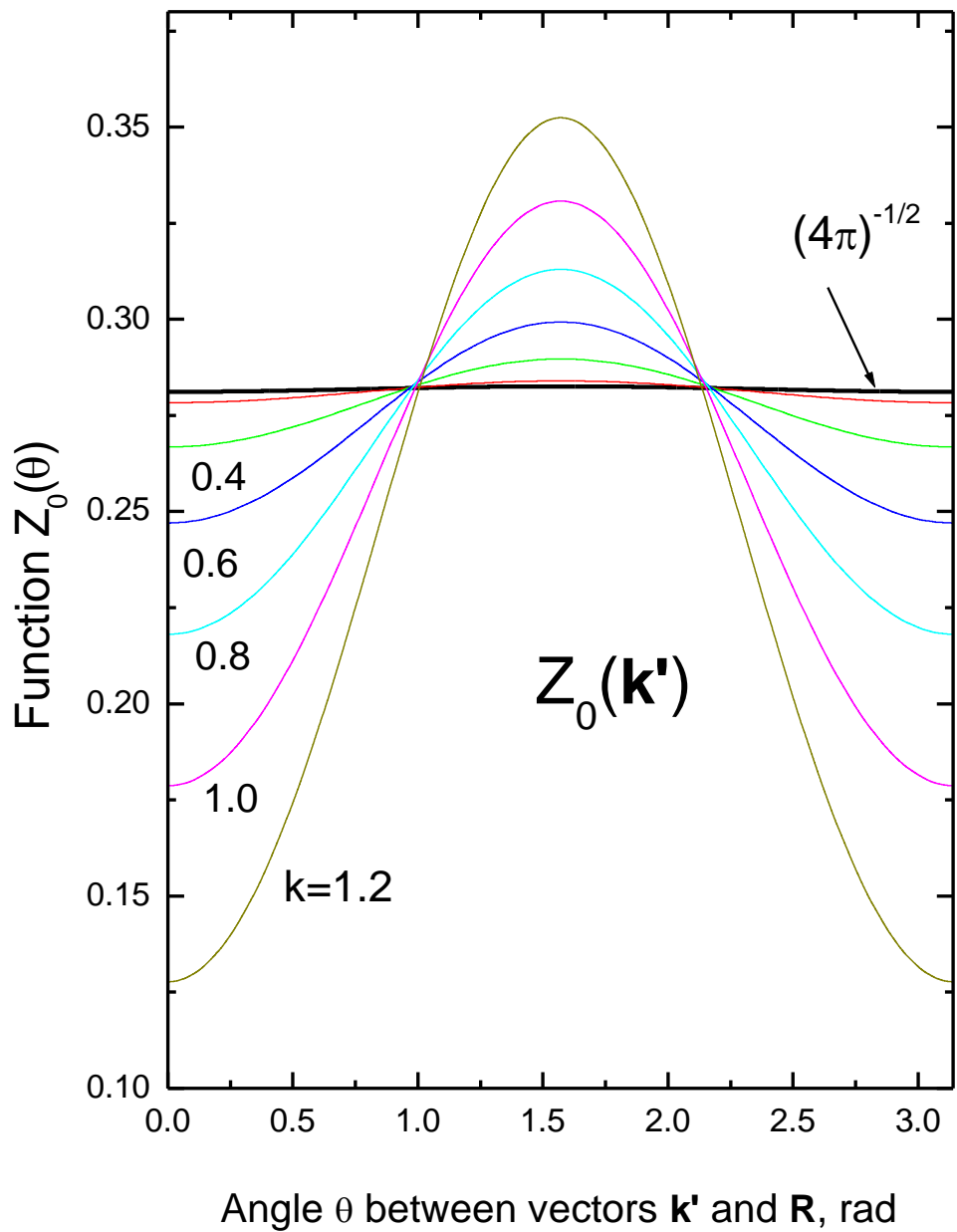

**Figure 2.** Function $Z_0(\mathbf{k})$ versus the angle $\theta$ between the vectors $\mathbf{k'} = k\,\mathbf{r}/r$ and $\mathbf{R}$. Function $Z_0(\mathbf{r})$ is represented by the same curves, and the angle $\theta$ is the angle between the vectors $\mathbf{r}$ and $\mathbf{R}$

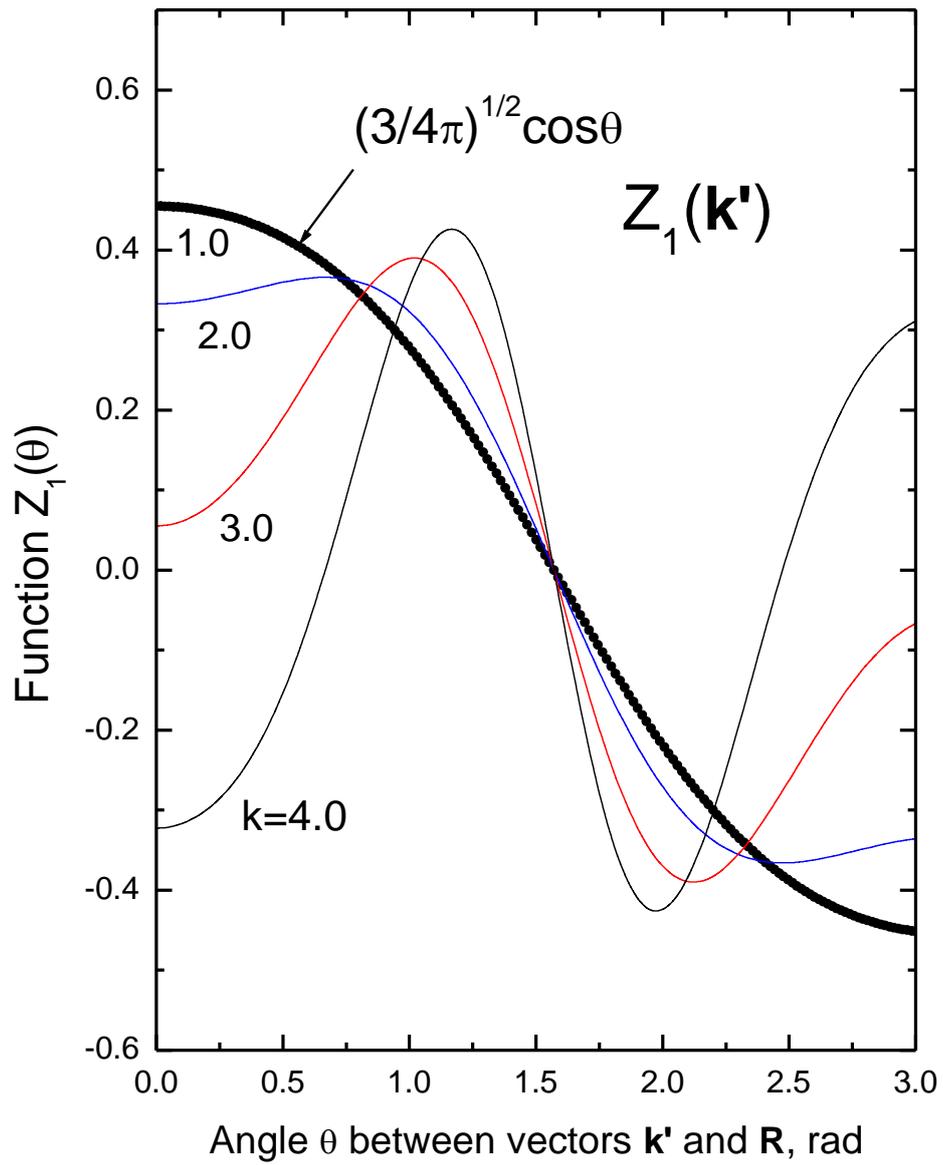

**Figure 3.** Function $Z_1(\mathbf{k'})$ versus the angle $\theta$ between the vectors $\mathbf{k'} = k\mathbf{r}/r$ and $\mathbf{R}$. Function $Z_1(\mathbf{r})$ is represented by the same curves, and the angle $\theta$ is the angle between the vectors $\mathbf{r}$ and $\mathbf{R}$.